**Material-Limited Switching in Nanoscale Ferroelectrics**

Tony Chiang,[1,†] John J. Plombon,[2] Megan K. Lenox,[3] Ian Mercer,[4]

Punyashloka Debashis,[2] Mahendra DC,[2] Susan Trolier-McKinstry,[4]

Jon-Paul Maria,[4] Jon F. Ihlefeld,[3,5] Ian A. Young,[2] and John T. Heron[1,6,†]

[1]Department of Materials Science and Engineering,

University of Michigan, Ann Arbor, MI 48109, USA

[2]Technology Research, Intel Corporation, Hillsboro, OR 97124, USA

[3]Department of Materials Science and Engineering,

University of Virginia, Charlottesville, VA 22904, USA

[4]Department of Materials Science and Engineering,

The Pennsylvania State University, University Park, PA 16802, USA

[5]Charles L. Brown Department of Electrical and Computer Engineering,

University of Virginia, Charlottesville, VA 22904, USA

[6]Applied Physics Program, University of Michigan, Ann Arbor, MI 48109, USA

[†] Corresponding author: chiangt@umich.edu, jtheron@umich.edu

**ABSTRACT**

The ferroelectric switching speed has been experimentally obfuscated by the interaction between the measurement circuit and the ferroelectric switching itself. This has prohibited the observation of real material responses at nanosecond timescales and lower. Here, fundamental polarization switching speeds in ferroelectric materials with the perovskite, fluorite, and wurtzite structures are reported. Upon lateral scaling of island capacitors from micron to nanoscales, a clear transition from circuit-limited switching to a material-limited switching regime is observed. In $La_{0.15}Bi_{0.85}FeO_3$ capacitors, switching is as fast as ~150 ps, the fastest switching time reported. For polycrystalline $Hf_{0.5}Zr_{0.5}O_2$ capacitors, a fundamental switching limit of ~210 ps is observed. Switching times for $Al_{0.92}B_{0.08}N$ are near 20 ns, limited by the coercive and breakdown electric fields. The activation field, instantaneous pseudo-resistivity, and energy-delay are reported in this material-limited regime. Lastly, a criterion for reaching the material-limited regime is provided. This regime enables observation of intrinsic material properties and favorable scaling trends for high-performance computing.

## MAIN

Ferroelectric memory elements have had a resurgence of interest due to the development of new families of materials with improved CMOS integrability, addressing the growing demand for energy-efficient and high-speed computing.[1–5] The electric field switchable polarization states make them promising for next-generation information and communication technologies.[6–8] One critical metric for this technology is its speed, often benchmarked by the delay, latency, or operations per second. Over the 100+ year history of ferroelectrics,[9] however, most reports on the 10-90% switching times involve micron- to cm-scale capacitors.[10–14] In this size range, a linear scaling of the switching time with capacitor area is observed[8,11,15] (Fig. 1a). The trend leads to a discrepancy between nucleation and growth models of ferroelectric switching and experimental observation. In nucleation and growth models, time is needed to nucleate a reversed domain, and additional time is needed for the boundaries to propagate through the material. One might then expect a saturation of the switching time when the area is sufficiently small. Further, the domain wall velocity is expected to be related to the acoustic phonon velocity, a few 1000s m/s,[16–18] which would correspond to picosecond scale switching times for ~ 30 μm$^2$ capacitors or smaller, assuming the conservative case of a single nuclei. Additionally, the high linear dielectric constant often found in ferroelectric materials, with the presence of a series resistance (~100-10000 Ω), leads to a *RC* time constant believed to limit the switching kinetics in this regime (often 1-100 ns range and above). The finite inductance of the cabling is fixed and does not contribute to the scaling behavior. This is commensurate with the linear relationship between the switching time and the capacitor area. The challenge has been understanding and mitigating the circuit influence as it convolves with polarization switching kinetics so that fundamentally materials-limited behavior can be elucidated. Here, new understanding of this trend is provided and regimes in which the ferroelectric switching time is governed by fundamental material behavior are determined. Disparate scaling trends are observed for the two regimes in terms of both fundamental physical parameters as well as technologically motivated parameters.

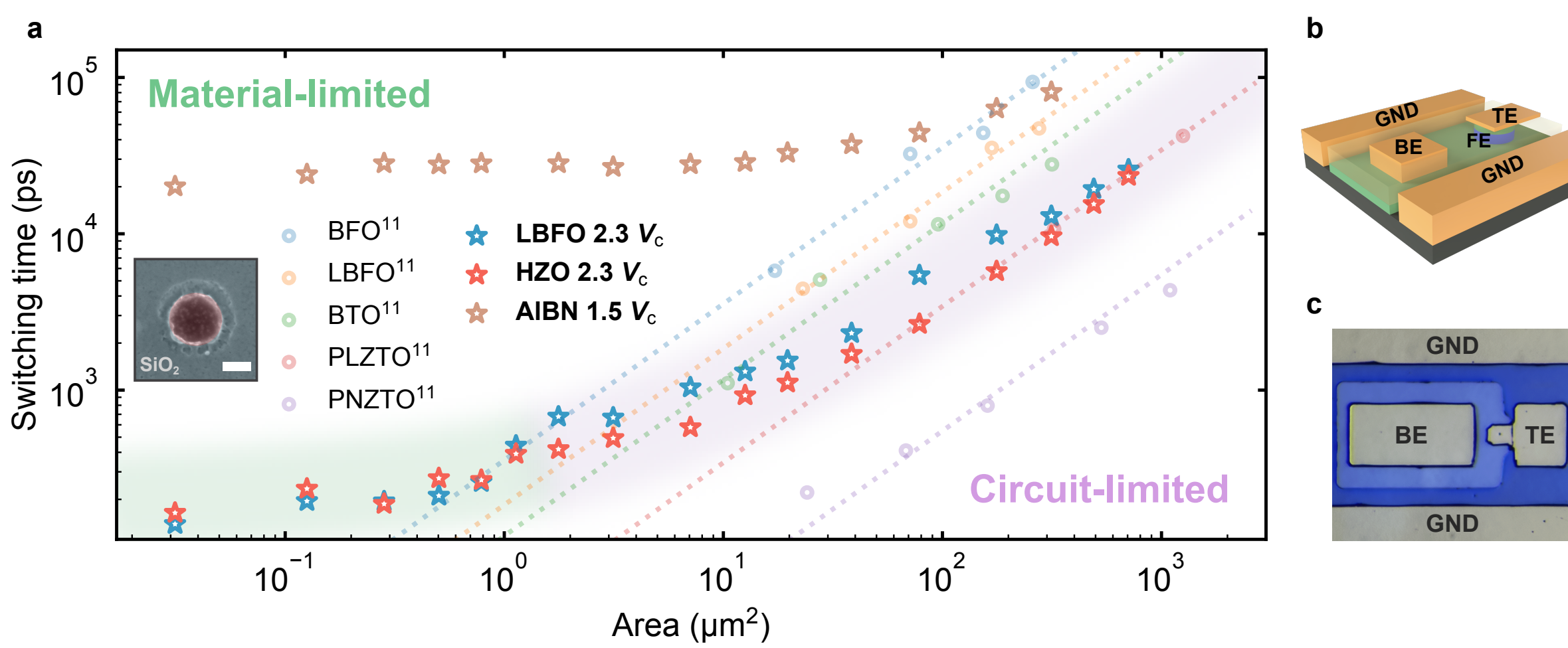

**Fig. 1 | Scaling behavior of switching time with ferroelectric capacitor area. a,** The switching time is defined as the time to go from 10-90% of full polarization reversal. For various materials reported in the literature (faint circular points) and those investigated here (red stars, HZO – $Hf_{0.5}Zr_{0.5}O_2$; blue stars, LBFO – $La_{0.15}Bi_{0.85}FeO_3$, and brown stars, AlBN – $Al_{0.92}B_{0.08}N$) are given. Voltages used for the materials in this study are given in terms of the coercive voltage ($V_c$) from ferroelectric hysteresis loops (obtained at 10-100 kHz, Fig. S1). At larger capacitor areas, the switching time is a linear function of the area, a signature of the circuit-limited regime. At small

capacitor areas, the switching time has a much weaker dependence on the capacitor area and reveals the material-limited regime where the switching is dominated by the physics of the material. In the smallest HZO and LBFO capacitors, the switching time is ~210 ps and ~150 ps at 2.3 $V_c$, respectively – the latter is the fastest switching time reported for any ferroelectric. Switching times for AlBN are near 20 ns, limited by the size of the electric field that can be applied without breakdown. Inset shows a SEM image of a 200 nm ferroelectric island (highlighted in red) surrounded by $SiO_2$. The scale bar is 100 nm. Data from the literature was obtained from Parsonnet et al.[11]. **b, c,** Schematic and the optical image of the ferroelectric capacitor semi-coplanar waveguides (MFM-sCPW) developed for this study, respectively. Ground (GND), back electrode (BE), and top electrode (TE) contacts are labeled.

To probe nanoscale capacitors at sub-nanosecond timescales, a semi-coplanar waveguide with an embedded vertical ferroelectric island capacitor, capable of transmitting a GHz signal with minimal parasitic capacitance, was designed for this study (Fig. 1b, 1c). Sample and measurement details can be found in the Methods section. Most importantly, the measurement configuration allows for simultaneous acquisition of the voltage across the ferroelectric capacitor (except in the case of the nitride ferroelectric due to the size of the switching voltage) and circuit current. In this study, three ferroelectric thin films are taken for comparison, polycrystalline $Hf_{0.5}Zr_{0.5}O_2$ (10 nm thick, HZO), epitaxial $La_{0.15}Bi_{0.85}FeO_3$ (20 nm thick, LBFO), and highly [001] textured $Al_{0.92}B_{0.08}N$ (15 nm thick, AlBN). The 10-90% switching time, determined from the polarization transients driven by a positive-up (PU) pulse train (down to 100 ps rise time) with the voltage setpoint indicated (relative to the quasi-static coercive voltage), for these three systems is shown in Fig. 1 as a function of the capacitor area. A clear non-linear scaling dependence is observed, and fast switching times (sub-ns) are realized with sub-micron diameter capacitors for both HZO and LBFO. These surpass the previously reported 10-90% switching times of 220 ps reported by Li, et al.[19] and 605 ps reported by Lyu, et al.[20] 200 nm diameter AlBN capacitors on the other hand switch at close to 20 ns, likely limited by the inability to source larger voltage pulses without experiencing significant leakage and dielectric breakdown with the much higher coercive field of AlBN.

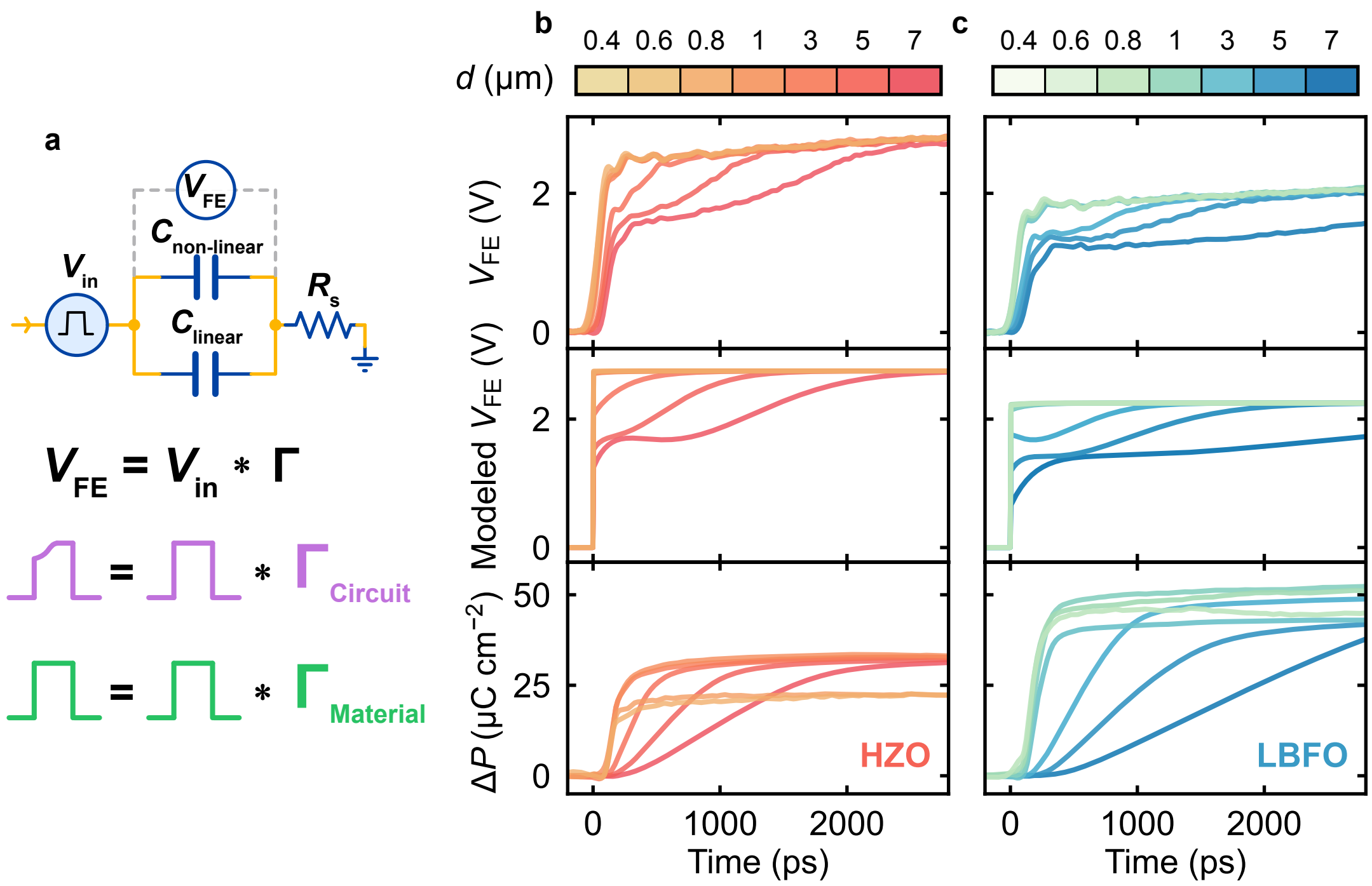

**Fig. 2 | Evolution of the capacitor voltage and current threshold vs capacitor diameter. a,** Schematic of the measurement circuit (top) without the 50 Ohm SMA transmission lines and fixed inductance from the strip line (bottom). Illustration of how the voltage pulse across the ferroelectric capacitor ($V_{FE}$) can be distorted, as it is the convolution of the square input voltage pulse and the circuit transfer function, Γ (which depends on the capacitor area, e.g. material and circuit-limited regimes). **b, c,** Experimentally measured voltage across the ferroelectric capacitor ($V_{FE}$) and the change in polarization (Δ*P*) as a function of capacitor area for HZO and LBFO, respectively. For large HZO and LBFO capacitors, $V_{FE}$ is distorted from square and does not reach the supplied voltage after an initial rise. After a temporary plateau, the voltage again rises and eventually reaches the applied value. The duration of the plateau scales with the capacitor area. In contrast to this, when capacitors are small enough, the ferroelectric capacitor voltage rises sharply to the applied value. The middle row panels show $V_{FE}$ modeled from a series resistor parallel capacitor-FE capacitor circuit and experimentally determined materials parameters from the polarization transients. Plateaus occur when a large current draw from the switching passes through the series resistor.

The switching can be categorized into two regions as shown in Fig. 1, the circuit-limited region where the switching time is dominated by a circuit effect (see Supplementary Information Section I) and a material-limited region where the fundamental switching kinetics of the material dominate. This is supported by the measured voltage transients ($V_{\mathrm{FE}}$) across the ferroelectric capacitor as a function of capacitor size, shown in Fig. 2. $V_{\mathrm{FE}}$ starts with a rapid rise, reaching the maximum (output setpoint) voltage rapidly for smaller capacitors. In contrast, larger capacitors show a rapid initial rise of the voltage which becomes sluggish before the setpoint voltage is reached, resulting in a significantly distorted pulse. This distortion becomes weaker as capacitor area decreases and completely disappears for HZO and LBFO capacitors of diameter around 1 µm. The significant deviation of $V_{\mathrm{FE}}$ from a step waveform in larger capacitors reveals the influence of the circuit. Evaluation of the $RC$ times from the linear dielectric component of the ferroelectric and background capacitances reveal that these are not significant to either the material- or circuit limited regime for our samples (Fig. S2).

This claim is supported by the circuit simulation of $V_{\mathrm{FE}}$ shown in the middle panels of Fig. 2b and c. Here, an $RC$ circuit model is constructed that incorporates a ferroelectric capacitor connected in parallel with a linear dielectric capacitor, representing the linear dielectric response of the ferroelectric plus a parasitic background, and connected to a series resistance. Ferroelectric switching is represented using the nucleation and growth process described by the Kolmogorov-Avrami-Ishibashi (KAI)[21–23] model using the expression:

$$\Delta P(t) = 2P_{\mathrm{r}}\left(1 - \exp\left[-\left(\frac{t}{t_0}\right)^n\right]\right)$$

where $P_{\mathrm{r}}$ is the remanent polarization, $t$ is time, $t_0$ is the characteristic time and $n$ is an exponent that includes nucleation but is often considered to be related to the dimensionality of the ferroelectric domain growth during switching. Here, $t_0$ and $n$ are fit parameters to the polarization transients shown in the bottom panels of Fig. 2. Using the fit parameters from the KAI equation, $V_{\mathrm{FE}}$ is modeled using a $RC$ circuit framework. A more detailed discussion of the model can be found in the Supplementary Information Section I. Notably, the model captures the sluggish rise of $V_{\mathrm{FE}}$ for large capacitors, which becomes less pronounced as the capacitor area decreases, consistent with the experimental $V_{\mathrm{FE}}$ data. The fast initial rise corresponds to the charging of the linear dielectric component ($C_{\mathrm{DE}}$). For large capacitors, there is a large ferroelectric switching current ($I_{FE}$) that follows the current from the ordinary dielectric. A large $I_{FE}$ leads to a significant voltage drop across the series resistor and thus a reduced or suppressed voltage across the ferroelectric capacitor, $V_{FE}$. Thus, the sluggish rise in $V_{FE}$ originates from the substantial demand for charge during ferroelectric switching. Notably, we find that the linear dependence of the switching time at large capacitor areas can be described in a model without any linear dielectric capacitor, which is appropriate when the relative permittivity is comparatively low (Fig. S3). As the capacitor area decreases, the total switching charge, and $I_{FE}$ induced, by the ferroelectric switching also decreases. The reduction in capacitor area then reduces circuit loading and the distortion of $V_{FE}$. The transition implies a cutoff voltage and $I_{FE}$.

$I_{FE}$ was also measured for all capacitor sizes and the maximum ($I_{FE,max}$) is plotted (multiplied by the series resistance and divided by the source input voltage) in Fig. 3a. Initially, $I_{FE,max}$ increases linearly with capacitor area and then deviates to an approximately saturated value for large capacitor areas. The delineation is consistent with the transition from material-limited to circuit-limited switching. In the linear region at small area, a KAI analysis of the ferroelectric switching current (which neglects leakage) describes the trend. Here,

$$I_{FE}(t) = A\frac{dP(t)}{dt} = \frac{2P_r A n}{t_0}\left(\frac{t}{t_0}\right)^{n-1} \exp\left[-\left(\frac{t}{t_0}\right)^{n}\right]$$

where the maximum ferroelectric switching current $I_{FE,max}$ occurring at time $t_{max}$ can be derived as:

$$I_{FE,max}(t_{max}) = \frac{2P_r A n}{t_0}\left(1-\frac{1}{n}\right)^{1-\frac{1}{n}} \exp\left[-\left(1-\frac{1}{n}\right)\right]$$

When the essential physics of the switching does not change with capacitor size, it would be expected that $t_0$ and $n$ do not change. In this case, the expression for $I_{FE,max}$ above has a linear relation to the capacitor area. Further, $t_0$ can be extracted and yields $132 \pm 3$ and $176 \pm 16$ ps for HZO and LBFO, respectively. In the circuit-limited regime, the saturation of the maximum current indicates that the current draw becomes clamped by the circuit. Thus, in the circuit-limited regime, the time required to supply sufficient charge for ferroelectric switching is a linear function of the ferroelectric capacitor area (total charge). The power limit of the supply unit was considered in the analysis (see Supplemental Section II). The power consumption for all capacitor sizes is well below the power limit of the supply. Therefore, the voltage and current transients unambiguously reveal the boundary between material-limited and circuit-limited switching regimes, with a crossover occurring at a capacitor diameter ~ 1 μm for the HZO and LBFO samples.

From Fig. 3a, a criterion is proposed for reaching the material-limited regime. This is the condition where the drop in supplied voltage ($V_{\text{in}}$) due to the ferroelectric current being drawn across the series resistance must remain below 10%. Thus, when the leakage and parasitic capacitance is small:

$$\Delta V_{\text{in}} = I_{\text{FE}} R_{\text{s}} \leq 0.1 V_{\text{in}}$$

Or when considering $I_{\text{FE,max}}$ in a KAI model

$$\frac{I_{\text{FE,max}} R_{\text{s}}}{V_{\text{in}}} = \frac{2 P_{\text{r}} A R_{\text{s}} n}{V_{\text{in}} t_0} \left(1 - \frac{1}{n}\right)^{1-\frac{1}{n}} \exp\left[-\left(1 - \frac{1}{n}\right)\right] \leq 0.1$$

which provides a condition for the ferroelectric capacitor area series resistance ($AR_{\text{s}}$) product as $t_0$ is roughly 100 ps and $n$ is commonly between 1 and 4 for many ferroelectrics. Thus, for $n$ = 3, $P_{\text{r}}$ = 40 μC cm$^{-2}$, $V_{\text{in}}$ = 1 V, and $t_0$ = 100 ps, $AR_{\text{s}} \leq$ 8.44 x 10$^{-8}$ Ω cm$^{2}$. For $R_{\text{s}}$ = 100 Ω, a critical diameter of around 3 μm would be expected. Note that the expression with $I_{\text{FE,max}}$ is in general voltage-dependent, which requires $t_0$ (and possibly $n$) to be voltage-dependent. The AlBN highlights the voltage dependence (via the relative size of the driving force for switching). Despite the large $P_{\text{r}}$, the reduced relative driving force (1.5 $V_{\text{c}}$ compared to 2.3 $V_{\text{c}}$ for HZO and LBFO) reduces $\frac{dP(t)}{dt}$ and the resulting current, such that most of the capacitor areas surveyed here are in the materials-limited regime. The proposed criterion establishes a guideline for practitioners to achieve the material-limited condition. The observation of the material-limited ferroelectric switching regime allows for determination of key ferroelectric switching parameters without the influence of the circuit.

The tracking of $V_{\text{FE}}$ along with the transient ferroelectric switching current, $I_{\text{FE}}$, enables determination of the energy delay characteristics, a figure of merit that represent the balance between speed and energy efficiency,[24–26] shown in Fig. 3b-c. Here we demonstrate favorable scaling of the energy-delay and femtojoule level operation (5 fJ for 200 nm diameter LBFO), with the energy consumption showing a markedly different dependence on capacitor area between the two regimes. The trade-off between energy and delay scales more favorably in the material limited regime, leading to anticipated sub-fJ dissipation at 40 nm diameter 2.25 V supply for LBFO (~60 aJ dissipation at 10 nm diameter) and offers significant advantages for logic applications. Power density versus delay, a critical metric for high-throughput computations (shown in Fig. 3d), reveals a favorable (decreasing power density and delay with decreasing capacitor area) scaling trend for both HZO and (more so for) LBFO, with a fit slope ~ -0.01 W s for LBFO. From the trends, the power density would cross 1 W mm$^{-2}$ at ~90 ps (at 2.25 V). Note that the power density appears larger than typical values because it is calculated based on the area of the ferroelectric capacitor, which is smaller than the overall circuit cell area.

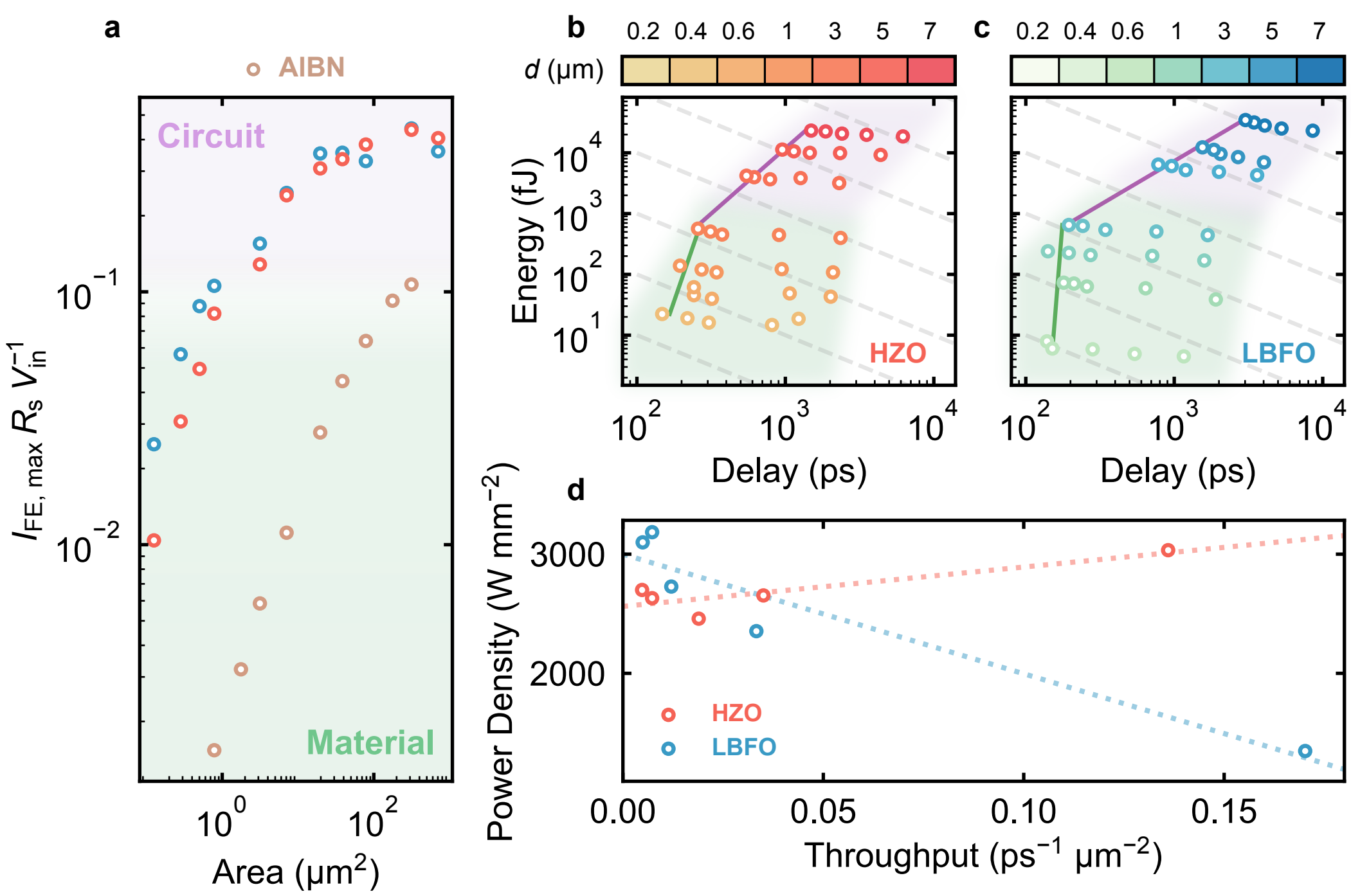

**Fig. 3 | Material-limited criterion and scaling of performance efficiency metrics. a,** Peak ferroelectric current, plotted as a ratio of voltage drop across a series resistor to the source voltage, as a function of capacitor area for HZO (red), LBFO (blue), and AlBN (brown) for the switching voltages explored. A plateau for large capacitors is observed that corresponds to the circuit-limited

regime (purple), and a material-limited regime (green) for smaller capacitors. The transition between these regions roughly occurs at 0.1 revealing a criterion for reaching the material-limited regime. **b, c,** Energy-delay contour for HZO (b) and LBFO (c) as a function of capacitor area and voltage. The grey dashed lines indicate constant energy-delay products, spaced by decade. Energy consumption exhibits strong dependence on capacitor area and can be traded with delay (by changing the driving voltage). The color shading highlights the circuit-limited regime (purple) and the material-limited regime (green), which exhibit distinct trends. The fits to these regimes are for fixed $V_{in}$ and reveal the different scaling trends. 5 fJ for 200 nm diameter LBFO was observed but extrapolation of the trend leads to sub-aJ at 40 nm diameter at the same drive voltage. **d,** Power density versus throughput for HZO and LBFO at various capacitor sizes. The dashed lines represent linear fits, revealing a more favorable energy-delay scaling trend in LBFO (2.25 V) compared to the weaker dependence observed in HZO (3 V).

In this direction, fundamental materials parameters in the material-limited regime were contrasted with the circuit-limited case. The voltage-dependent switching time across these two regimes was fit to a semi-empirical model (i.e. Merz's law:[27] $t_{\text{switch}} = t_{\infty} \exp\left(\frac{E_a}{E}\right)$ ) that describes the switching time as a function of the material dependent activation field ($E_a$) and the applied electric field ($E$). Fig. 4a shows the determined $E_a$, normalized by the quasi-static coercive field ($E_c$). Again, two regions are observed, consistent with those observed in Fig. 1. In the circuit-limited region, the apparent activation field increases linearly with decreasing capacitor size, while in the material-limited regime, it flattens to a roughly constant value. For ferroelectrics in these size ranges where no critical physics changes (e.g. many nuclei form and grow, a many-grain structure persists across all HZO capacitor sizes – Fig. S9), a constant $E_a$ would be expected.[28] This is, in fact, observed in the material-limited regime. The linear dependence in the circuit-limited regime is a consequence of the reduced time-averaged electric field. This clearly reveals that the conventional use of Merz law in the circuit-limited regime is inappropriate as the electric field is evolving on time scales characteristic of the ferroelectric switching and that measuring the capacitor voltage is necessary to elucidate material parameters regarding ferroelectric switching dynamics. Note that in our prior study, we examined the electric field distribution in a pillar capacitor structure for HZO with an identical device structure using finite element analysis where a modest field amplification was observed within 5 nm of the edge.[29] This supports excluding the influence of inhomogeneous electric field distribution effects in the scaling behavior.

Further, the measurement of $V_{FE}$ and $I_{FE}$ allows the transient pseudo-resistivities ($\rho_{pseudo} = \frac{V_{FE}}{I_{FE}}\frac{A}{t}$) (Fig. 4b, 4c) to be determined. Remarkably, the instantaneous pseudo-resistivity of LBFO and HZO drops by 8 orders of magnitude, to below 10 Ω cm, during switching with the time of the effective resistivity minima matching that of $I_{FE,max}$. Due to the circuit-limited effects on $V_{FE}$ and $I_{FE}$ at large capacitor area, the minimum pseudo-resistivity also shows two distinct switching regimes. The differential pseudo-resistivity is negative (positive) at initial (later) times and spans nine orders of magnitude. These open doors to highly nonlinear circuits in the time domain at picoseconds timescales in ferroelectric materials.

A 400 nm HZO capacitor was used to evaluate the fundamental switching limit of a multi-grain capacitor. The Merz behavior was examined up to 3.75 MV cm$^{-1}$ (2.88 $E_c$) as shown in Fig. 4d. The switching time saturates around 2.75 MV cm$^{-1}$ (2.12 $E_c$) and remains nearly constant with higher voltages (211 ± 11 ps). A comparison of the switching time with the $V_{FE}$ rise time and the time constant $RC_{non\text{-}switch}$ of the non-switching pulse shows that the switching time is two times slower than either these two timescales, revealing that the switching time is not limited by the circuit or electronics. The speed limit for the many-grain HZO capacitor may arise from the suppression of the domain wall motion via grain boundary pinning,[30] or the likely complex domain and possible multiphase structures,[31,32] commensurate with the nucleation limited switching picture.[33,34] One can then expect that a device that harnesses the full-scale ferroelectric switching speed of many-grain HZO thin films can operate up to 5 GHz, greatly surpassing other emerging non-volatile memories,[1,26,35–37] and demonstrating potential for both memory and logic applications. While we were unable to observe a clear saturation, the time constant from the Merz analysis of LBFO capacitors suggests a switching speed limit of 147 ± 49 ps.

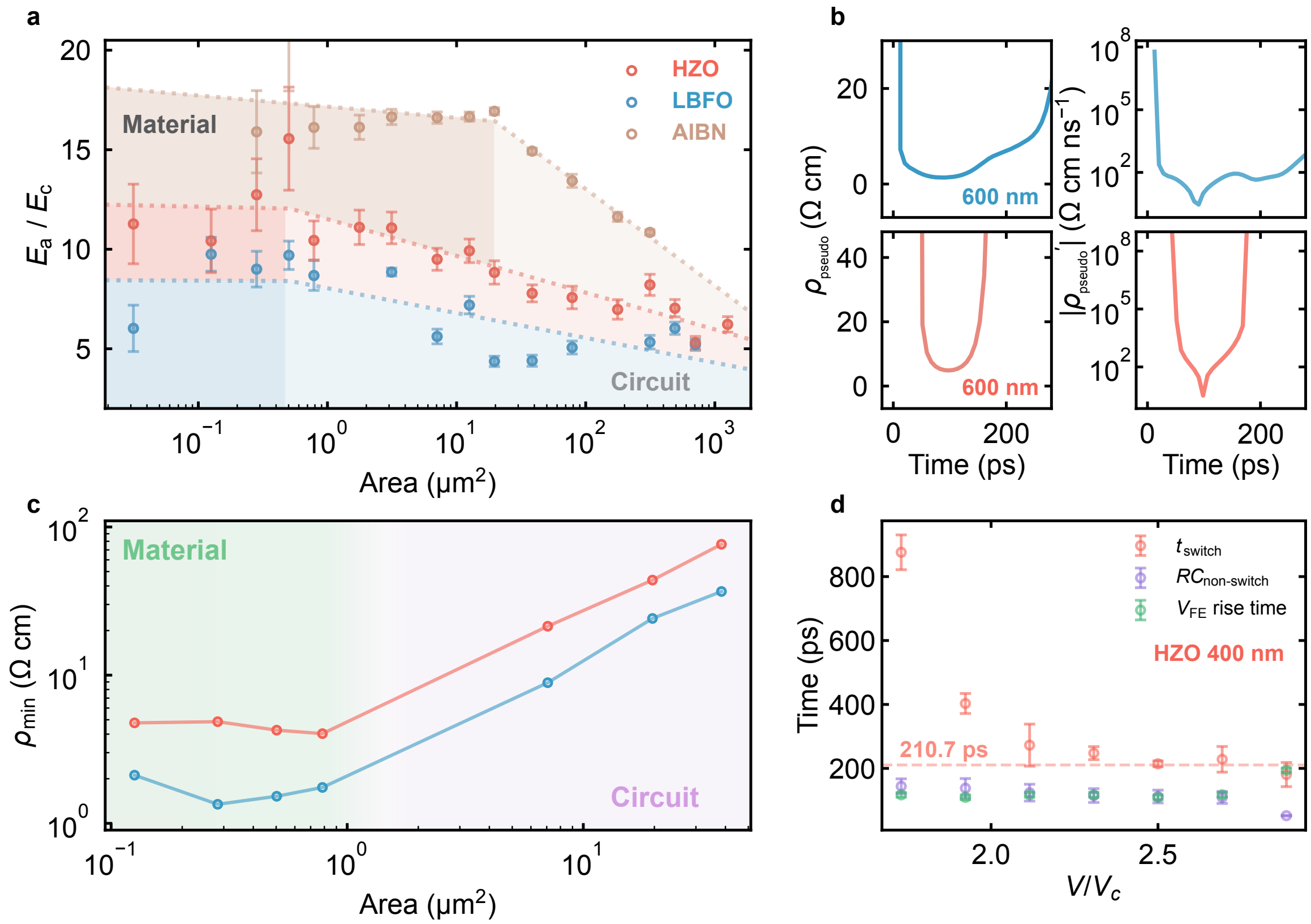

**Fig. 4 | Scaling trends of the Merz activation field, instantaneous pseudo-resistivity, and HZO switching limit. a,** Activation field as a function of capacitor area for the three ferroelectrics considered. A material-limited regime is observed where $E_a$ stays relatively constant and a circuit-limited regime that leads to a linear change in $E_a$ with capacitor area. **b,** Transient pseudo-resistivity ($\rho_{\text{pseudo}}$) and the absolute value of its derivative ($|\rho_{\text{pseudo}}{}'|$) for 600 nm LBFO (blue) and HZO (red) capacitors, demonstrating a decrease of more than eight orders of magnitude in pseudo-resistivity from the DC value. **c,** Minimum transient pseudo-resistivity (defined as $\frac{V_{\text{FE}}}{I_{\text{FE}}}$) of LBFO and HZO capacitors during ferroelectric switching as a function of area where again the two regimes are seen. **d,** 10-90% switching time of a 400 nm polycrystalline HZO capacitor as a function of input voltage, plotted along with the circuit *RC* time constant extracted from the non-switching current and the rise time of the voltage pulse. A saturation of the switching time at ~210 ps is observed.

In summary, material-limited ferroelectric switching is demonstrated in nanoscale ferroelectric capacitors. In this regime, fundamental materials parameters can be determined. The regime is most clearly distinguished by the absence of voltage droop during switching. We also propose a threshold criterion for the ferroelectric switching current for ease in regime evaluation and circuit design. We report crossovers in areal scaling behavior for several physical parameters, revealing where the circuit influence is no longer dominant. In this regime, the fastest switching observed is 147 ± 49 ps for a 200 nm LBFO capacitor, and a switching limit of 211 ± 11 ps for many-grain HZO. Finally, the energy-delay and power dissipation metrics are examined in the material-limited regime where a more favorable scaling trend with capacitor area is observed relative to the circuit-limited regime where the current is clamped. With nanoscale capacitors, femtojoule operation with few hundred picosecond delay is demonstrated, revealing promising outlook for applications. These results broaden the fundamental understanding of ferroelectric switching, facilitate the design of ferroelectronic devices, and demonstrate the feasibility of multi-GHz operations.

## METHODS

### Thin Film Growth

The epitaxial 15% La-doped $BiFeO_3$ ($SrRuO_3$ 20 nm/LBFO 20 nm/$SrRuO_3$ 50 nm) was grown by pulsed laser deposition with a (110)-oriented $DyScO_3$ single crystal serving as the substrate. The $Hf_{0.5}Zr_{0.5}O_2$ (W 20 nm/HZO 10 nm/W 50 nm) film was grown using plasma-enhanced atomic layer deposition on an undoped (001)-oriented silicon substrate, with the bottom and top tungsten layers deposited by DC magnetron sputtering.[38] The film was crystallized at 600 °C for 30 seconds via rapid thermal annealing. The microstructure comprised grains that span the thickness of the film and have lateral dimensions of 22.9 nm, on average. The $Al_{0.92}B_{0.08}N$ (W 20 nm/AlBN 15 nm/Ti 5 nm/Ti 90 nm) film was prepared using magnetron sputtering on an undoped (001)-oriented silicon substrate.[5]

### MFM-sCPW Fabrication

The ferroelectric capacitor semi-coplanar waveguides (MFM-sCPW) were patterned with a series of E-beam and optical photolithography steps. Starting from the first layer, capacitors were defined with a negative ma-N 2405 E-beam resist spun at 5000 rpm with a softbake at 100 °C for 4 minutes, then exposed using a JEOL JBX-6300FS electron beam lithography system. The sample was then ion milled at a 25° tilt with the negative resist serving as an etch mask, to the bottom electrode to create self-aligned top electrode / ferroelectric islands. The samples were then backfilled with 140 nm of $SiO_2$ using magnetron sputtering (PVD 75 Proline, Lesker) for low κ isolation between the top and bottom electrodes. After the deposition, the E-beam resist was removed by ultrasonicating in Remover PG. Next, a series of photolithography steps were performed to define patterns to etch away the surrounding leftover bottom electrodes and pattern the ground planes. Finally, gold (300 nm)/titanium (10 nm) top contacts were deposited with an E-beam evaporator.

**Ultra-fast Large Signal Ferroelectric Switching Measurements**

Ultra-fast large signal measurements were performed in an RF probe station, with the top and bottom electrodes contacted by the Ground-Signal-Ground probe (Model 40A, GGB) with a pitch of 100 μm. Low loss RF cables (SF526S/11PC35, HUBER+SUHNER) with a total length of 3500 mm were used for signal transmission. The voltage pulse train was generated by a pulse generator with a rise time down to 100 ps, connected to the top electrode, while the signal was recorded with a 13 GHz oscilloscope with a 50 Ω input, connected to the bottom electrode. With this setup, the current in the series *RC* circuit can be determined using the oscilloscope 50 Ω internal loading. For real-time voltage monitoring, the high-impedance probes were landed at the top and bottom electrodes to capture the differential voltage across the capacitor. All measurements were conducted in real-time with averaging over four single-shot measurements. The positive-up (PU) pulse train with a pulse width of 1 μs was used to probe ferroelectric switching transients to subtract the non-ferroelectric contribution, with a reset pulse with negative polarity applied before the PU pulse train. Both the LBFO and HZO were measured at room temperature, and AlBN was measured at 130 °C to reduce the coercive field.[39] A DC bias up to 1.2 V for compensating the imprint field was added for LBFO during the high-speed measurement to prevent back-switching.

## DATA AVAILABILITY

All data that support the findings of this study are available from the corresponding author on reasonable request.

## ACKNOWLEDGMENTS

T. C. and J. T. H. acknowledge support from the Intel FEINMAN 2.0 program. The authors acknowledge the University of Michigan College of Engineering for financial support and the Michigan Center for Materials Characterization for use of the instruments and staff assistance. This work was performed in part at the University of Michigan Lurie Nanofabrication Facility. HZO films were prepared under support from the Laboratory Directed Research and Development Program at Sandia National Laboratories. Sandia is a multimission laboratory managed and operated by National Technology and Engineering Solutions of Sandia LLC, a wholly owned subsidiary of Honeywell International Inc. for the U.S. Department of Energy's National Nuclear Security Administration under Contract No. DE-NA0003525. I.M., S.T-M., and J-P. M. acknowledge support from the Center for 3D Ferroelectric Microelectronics Manufacturing, an Energy Frontier Research Center funded by the U.S. Department of Energy, Office of Science, Basic Energy Sciences under Award # DE-SC0021118 for the growth of the wurtzite ferroelectric films, and for interpretation of the switching kinetics.

**Supplementary Information for**

# Material-Limited Switching in Nanoscale Ferroelectrics

Tony Chiang,[1, †] John J. Plombon,[2] Megan K. Lenox,[3] Ian Mercer,[4]

Punyashloka Debashis,[2] Mahendra DC,[2] Susan Trolier-McKinstry,[4]

Jon-Paul Maria,[4] Jon F. Ihlefeld,[3, 5] Ian A. Young,[2] and John T. Heron[1, 6, †]

[1]Department of Materials Science and Engineering,

University of Michigan, Ann Arbor, MI 48109, USA

[2]Technology Research, Intel Corporation, Hillsboro, OR 97124, USA

[3]Department of Materials Science and Engineering,

University of Virginia, Charlottesville, VA 22904, USA

[4]Department of Materials Science and Engineering,

The Pennsylvania State University, University Park, PA 16802, USA

[5]Charles L. Brown Department of Electrical and Computer Engineering,

University of Virginia, Charlottesville, VA 22904, USA

[6]Applied Physics Program, University of Michigan, Ann Arbor, MI 48109, USA

[†] Corresponding authors: chiangt@umich.edu, jtheron@umich.edu

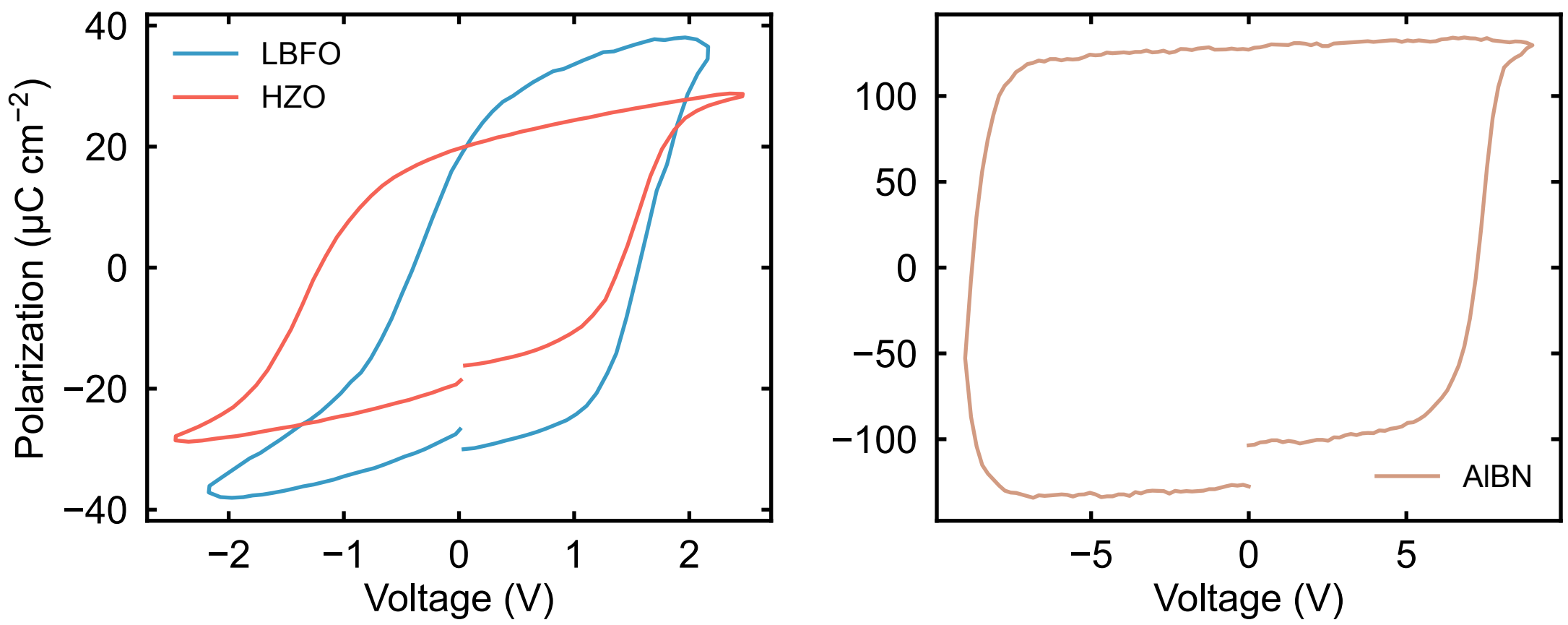

**Fig. S1 | Ferroelectric hysteresis.** Quasi-static (100 kHz for LBFO and HZO, 10 kHz for AlBN) ferroelectric hysteresis measurement with a 20 μm diameter capacitor for LBFO (blue), HZO (red), and AlBN (brown).

## I. Support Evidence for Circuit Limited Switching

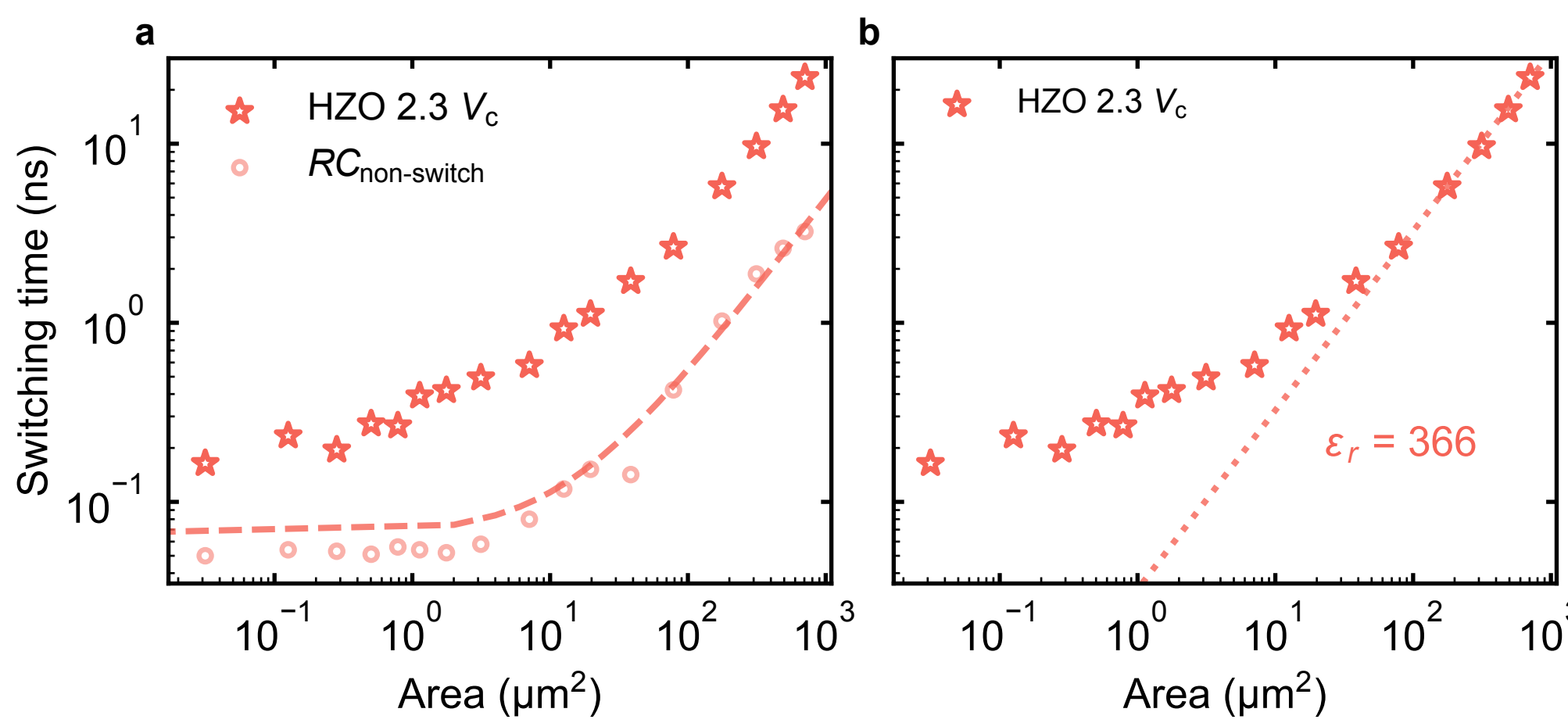

**Fig. S2 | Circuit-limiting effect on ferroelectric switching times.** The stars show experimental data of the 10-90% switching time for the HZO sample and the hollow circles (**a**) representing *RC* time constant of the non-switching pulse which contains the capacitance from the linear dielectric component of the ferroelectric and the parasitic background. The dashed line in **a** is the linear fit of the *RC* time constant, with an y-intercept of 65 ps revealing the level of the parasitic capacitance; this highlights the fast circuit response. The dotted line in **b** is a linear fit to the large (circuit-limited) area capacitor data and yields a relative permittivity of 366, an order of magnitude higher than the linear relative permittivity of HZO. This reveals that the circuit-limiting effect for these samples is not related to a linear circuit capacitance. This applies in cases where the switching current dominates over the dielectric contribution.

The transient voltage ($V_{\mathrm{FE}}$) across the ferroelectric capacitor is modeled using an $RC$ circuit, consisting of two ideal (as the leakage from the experimental data is negligible on the picosecond to nanosecond timescale) capacitors connected in parallel, one representing the non-linear ferroelectric component ($C_{\mathrm{FE,NL}}$) and the other representing the linear ferroelectric component and the parasitic background from electrodes/ test structure geometry ($C_{\mathrm{PB}}$). The parasitic component can be described by a standard linear capacitor and the transient current would be:

$$C_{\mathrm{PB}} = \varepsilon_0 \varepsilon_r' \frac{A}{d}$$

$$I_{\mathrm{PB}}(t) = \frac{V_{\mathrm{in}}}{R_{\mathrm{s}}} \exp\left[-\frac{t}{R_{\mathrm{s}} C_{\mathrm{PB}}}\right]$$

where $\varepsilon_0, \varepsilon_r', A, d, V_{\mathrm{in}}$, $R_s$ are the vacuum permittivity, effective relative permittivity of the linear background signals, capacitor area, ferroelectric film thickness, applied voltage, and series resistance, respectively.

The circuit can be split into two stages in time. First, the dielectric charging that occurs initially upon voltage bias and the ferroelectric component is not yet active. Here the circuit can be treated as a single dielectric capacitor. After some time delay, the ferroelectric switching is active. The ferroelectric switching dynamics described by the KAI model yield an evolving polarization and current given by:

$$P(t) = 2P_{\mathrm{r}}\left(1 - \exp\left[-\left(\frac{t}{t_0}\right)^n\right]\right)$$

$$I_{\mathrm{FE}}(t) = A\frac{dP(t)}{dt}$$

The polarization transients can be accurately described using the experimental fit parameters for $t_0$ and $n$. $A$ represents the area of the ferroelectric capacitor.

The total current through the circuit is then given by the sum of the linear dielectric charging current and the ferroelectric switching current ($I_{\mathrm{FE}}$):

$$I_{\mathrm{tot}} = I_{\mathrm{PB}} + I_{\mathrm{FE}}$$

Finally, the transient voltage across the ferroelectric capacitor is given by:

$$V_{\mathrm{FE}} = V_{\mathrm{in}} - R_{\mathrm{s}} I_{\mathrm{tot}}$$

From this equation, it is evident that a large $I_{\mathrm{tot}}$ reduces the transient voltage $V_{\mathrm{FE}}$, which can be attributed to a rapid growth in $I_{\mathrm{FE}}$ during ferroelectric switching. This effect is particularly pronounced in larger devices with greater device area $A$, causing the in $V_{\mathrm{FE}}$ dependent on device size. When the capacitor area is large and $I_{\mathrm{PB}}$ is negligible, the total current is dominated by the ferroelectric switching current. This clearly illustrates the interplay between the voltage across the ferroelectric capacitor and the switching of the ferroelectric. Taking the time average of the equation yields:

$$\langle V_{\mathrm{FE}} \rangle = V_{\mathrm{in}} - R_{\mathrm{s}} A \left[ \frac{(0.8)\ 2P_r}{t_{switch}} \right]$$

Here, $\langle V_{\mathrm{FE}} \rangle$ is the time averaged (across the 10-90% switching time or extracted from Merz analysis) voltage across the ferroelectric capacitor and $\frac{(0.8)\ 2P_r}{t_{switch}}$ represents the linearized switching using the 10-90% polarization switching time ($t_{switch}$). Rearranging this yields:

$$t_{switch} = \frac{1.6 R_{\mathrm{s}} A P_r}{V_{\mathrm{in}} - \langle V_{\mathrm{FE}} \rangle}$$

The expression reveals the linear area dependence of the switching time at large capacitor area. Note that $\langle V_{\mathrm{FE}} \rangle \sim \frac{1}{2} V_{\mathrm{in}}$ for large capacitor areas. Again, the physics is devoid of any linear dielectric phenomena. Figure S3 below plots the expression for $t_{\mathrm{switch}}$ above (using the time averaged (over the time for 10-90% switching) voltage) along with the experimental data from the HZO sample.

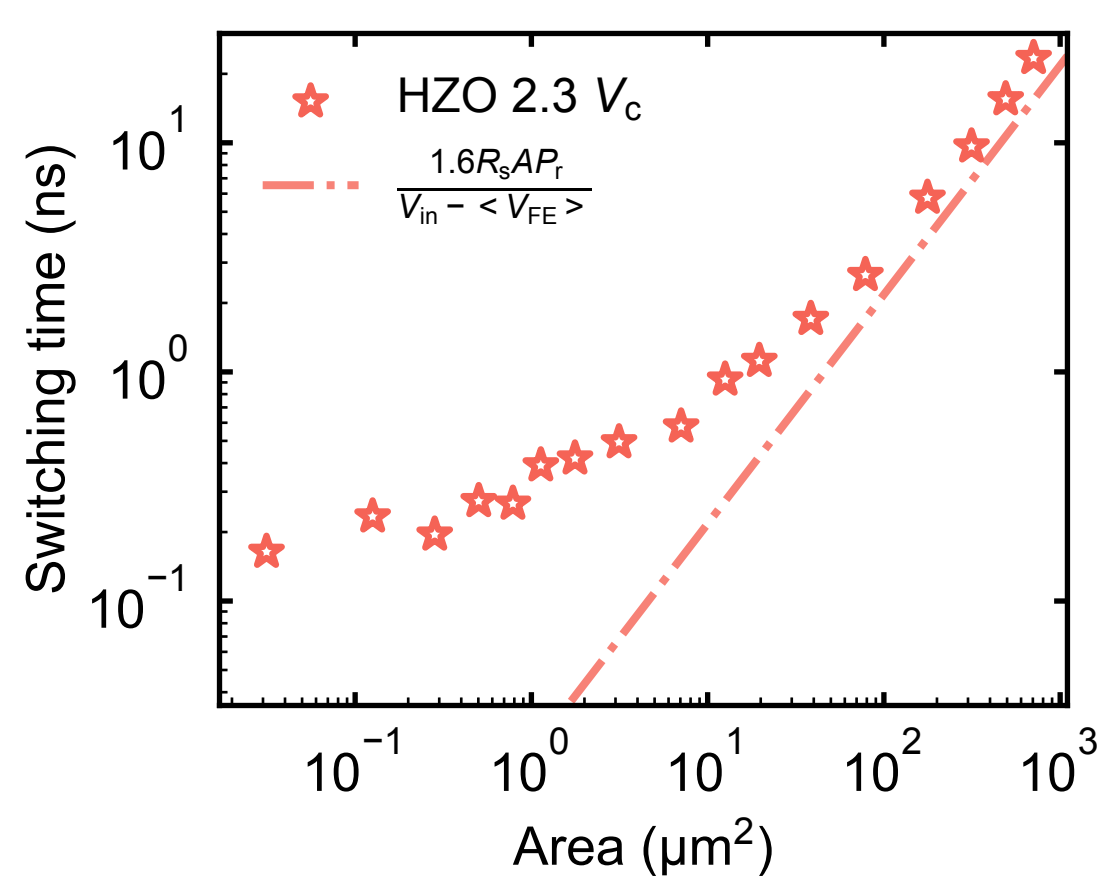

**Fig. S3 | Analysis of the 10-90% switching time for large capacitor areas.** Data is plotted in star and the dash dotted line is plot of expression of $t_{\mathrm{switch}}$ defined above.

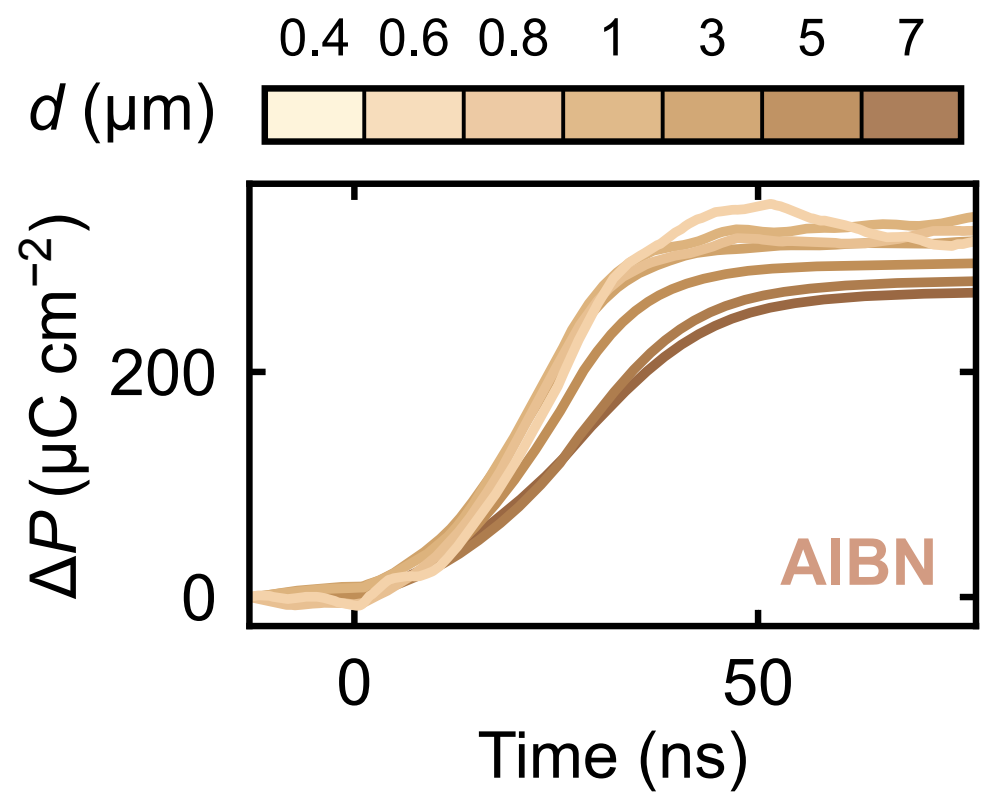

**Fig. S4 | Polarization vs time for AlBN capacitors.** The change in polarization ($\Delta P$) as a function of capacitor diameter for AlBN.

## II. Evaluation of Power Limitation

The power limit of the source could influence ferroelectric switching when the power demand exceeds this limit. Our source is designed to be connected to a minimum load of 50 Ω. This defines the maximum power that can be supplied. To show that our measurement results in both the material-limited and circuit-limited regime are not merely limited by the power limitation of the source, the peak ferroelectric switching current and the maximum power consumption as a function of capacitor diameter is considered (Fig. S5). With a 2.25 V (3 V, for HZO) output, the maximum current that can be supplied by the waveform generator would be 2.25 V / 50 Ω, which corresponds to 45 mA (60 mA). In Fig. S5a, we show that the peak current during the ferroelectric switching of both materials is much below this limit. Another way to examine this would be the power consumption. The maximum power that can be supplied by the source is $\frac{V^2}{R}$. Using a set voltage of 2.25 V (LBFO) and 3 V (HZO), and a minimum load of 50 Ω, this yields a maximum power supply of 101.25 mW (180 mW). Using capacitors with diameters of 7 μm and 1 μm, in Fig. S5b to Fig. S5e the power consumption ($V_{\mathrm{FE}}I_{\mathrm{FE}}$) is below this limit throughout the switching in both regimes.

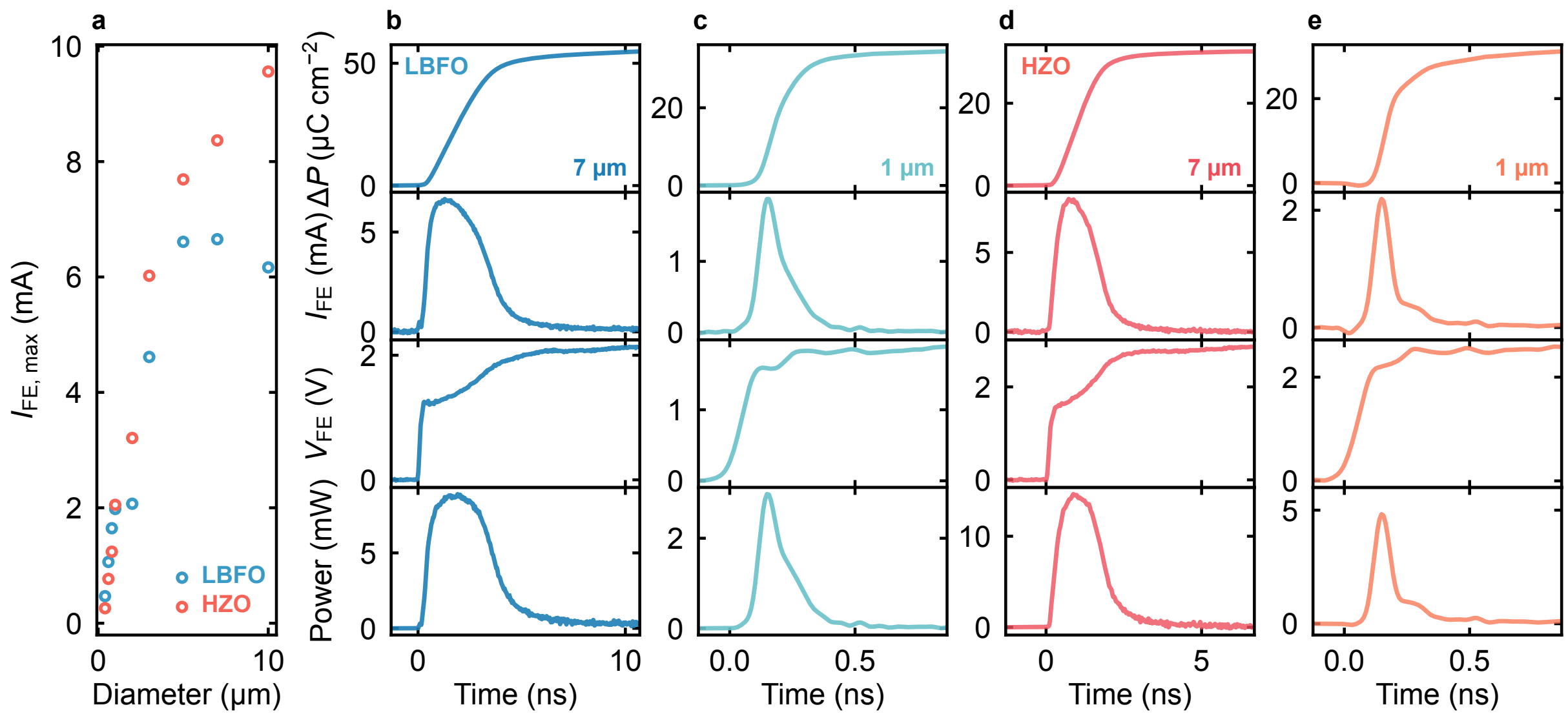

**Fig. S5 | Maximum current and power consumption of ferroelectric switching. a**, Peak ferroelectric switching current as a function of capacitor diameter from 2.25 V supply voltage for LBFO and 3 V for HZO. **b**, **c**, **d**, **e**, Transient polarization, ferroelectric switching current, voltage, and power consumption of 7 µm and 1 µm LBFO (blue) and HZO (red) capacitors.

## III. Extracting Material Parameters Using $I_{\mathrm{FE,max}}$

Starting with the KAI expression,

$$P(t) = 2P_{\mathrm{r}}\left(1 - \exp\left[-\left(\frac{t}{t_0}\right)^n\right]\right)$$

It follows that the current from the ferroelectric switch is:

$$I_{\mathrm{FE}}(t) = A\frac{dP(t)}{dt} = \frac{2P_{\mathrm{r}}An}{t_0}\left(\frac{t}{t_0}\right)^{n-1} \exp\left[-\left(\frac{t}{t_0}\right)^n\right]$$

where the maximum of $I_{\mathrm{FE}}$ can be determined by solving:

$$\frac{d}{dt}I_{\mathrm{FE}}(t) = \frac{d}{dt}\frac{2P_{\mathrm{r}}An}{t_0}\left(\frac{t}{t_0}\right)^{n-1} \exp\left[-\left(\frac{t}{t_0}\right)^n\right]\Bigg|_{t=t_{\mathrm{max}}} = 0$$

This leads to the following expression for $t_{\mathrm{max}}$ :

$$t_{\mathrm{max}} = t_0\left(\frac{n-1}{n}\right)^{1/n}$$

Thus, the maximum current $I_{\mathrm{FE,max}}$ has the form

$$I_{\mathrm{FE,max}}(t_{\max}) = \frac{2P_{\mathrm{r}}An}{t_0}\left(1-\frac{1}{n}\right)^{1-\frac{1}{n}} \exp\left[-\left(1-\frac{1}{n}\right)\right]$$

Note that $n$ must be greater than 1 and is often attributed to the dimensionality of the growth of the nuclei plus 1.[1–3] These expressions reveal a dependence of $I_{\mathrm{FE,max}}$ on $A$, as $n$ and $t_0$ are material parameters that will not vary with area if the essential physics does not change and if $V_{\mathrm{FE}}$ is not circuit-limited. As $1 < n \leq 4$ is often found, the $n$ terms are small and have little variance. Thus, the slope of $I_{\mathrm{FE,max}}$ vs $2P_{\mathrm{r}}A$ will have a slope $\sim 1/t_0$. $I_{\mathrm{FE,max}}$ vs capacitor area reveals two distinctive regimes (shown in Fig. S6), where a linear region is observed for smaller capacitors in the material-limited regime and a change of slope with further increase of capacitor size entering the circuit-limited regime due to the circuit influence on $t_0$ and $n$. Evaluating the slope in the material-limited regime leads to a $t_0$ of $176 \pm 16$ ps for LBFO, $132 \pm 3$ ps for HZO, and $22.3 \pm 0.9$ ns for AlBN.

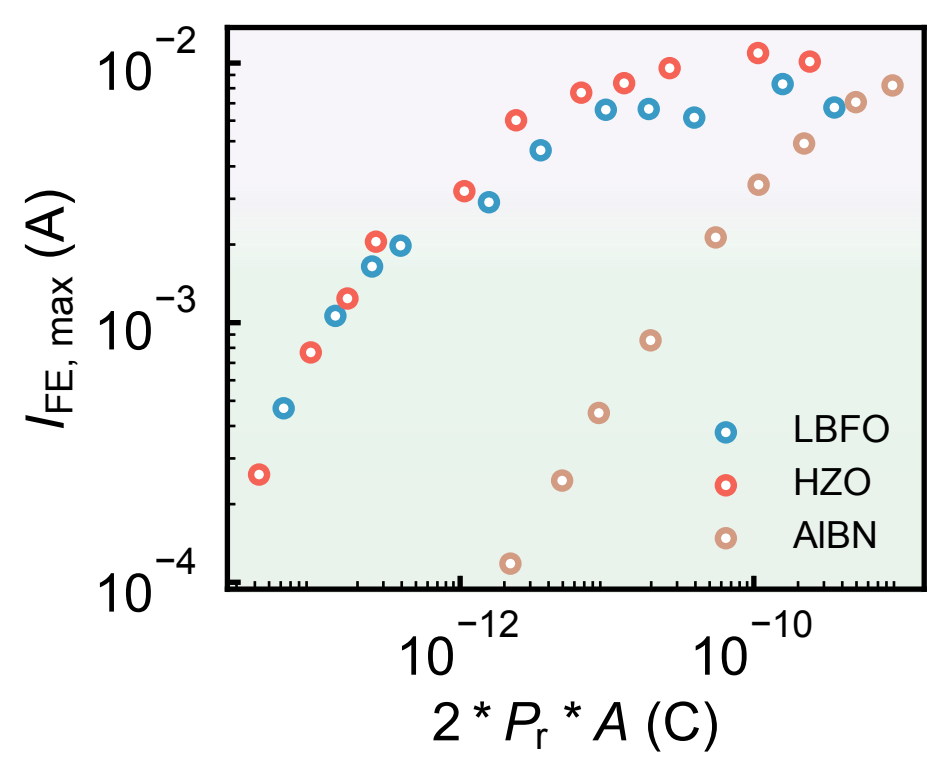

**Fig. S6 | Demonstration of the material-limited and circuit-limited regime using the maximum ferroelectric switching current.** The purple shaded region is the circuit-limited regime while that in green is the material limited regime for the field ranges explored. The boundary is around 2 mA for all samples.

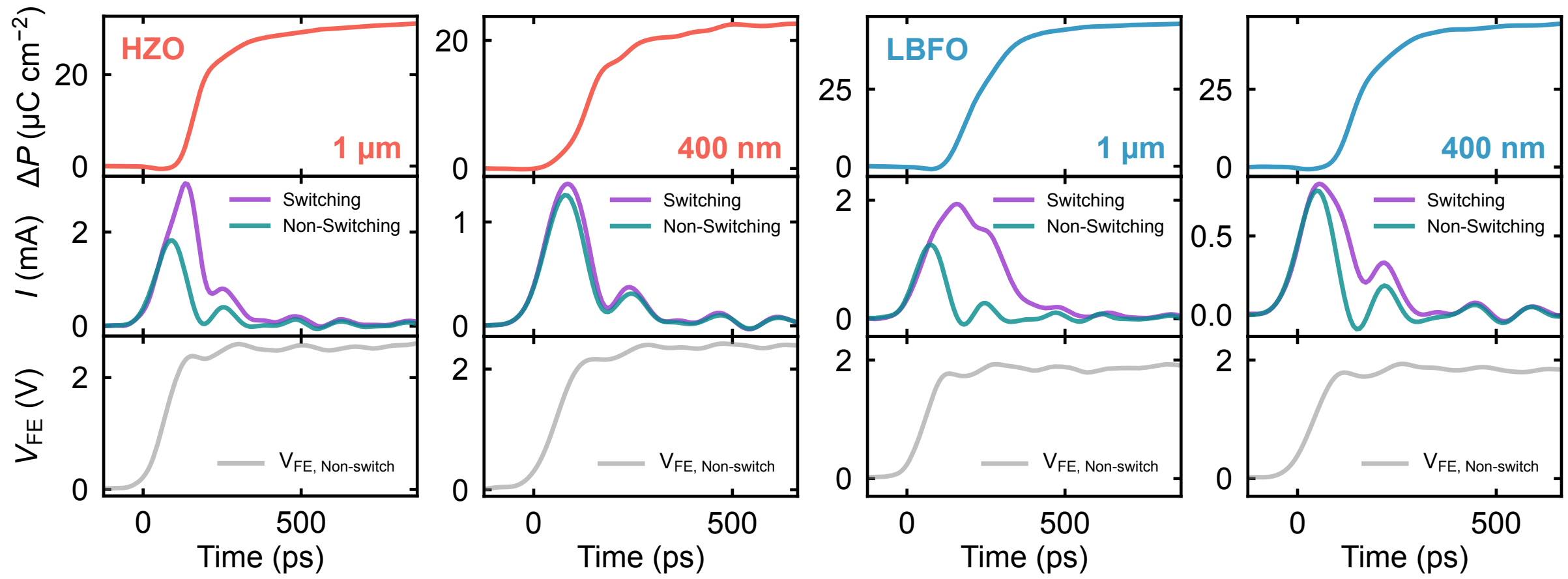

**Fig. S7 | Verification of high-speed measurement signal integrity.** The top panels show the polarization transient of 1 μm and 400 nm HZO and LBFO device. The center panels show the switching and non-switching currents, highlighting a clear ferroelectric switching behavior and the negligible leakage current at this timescale. The bottom panels show $V_{\mathrm{FE}}$ for the non-switching part, demonstrating the fast circuit response.

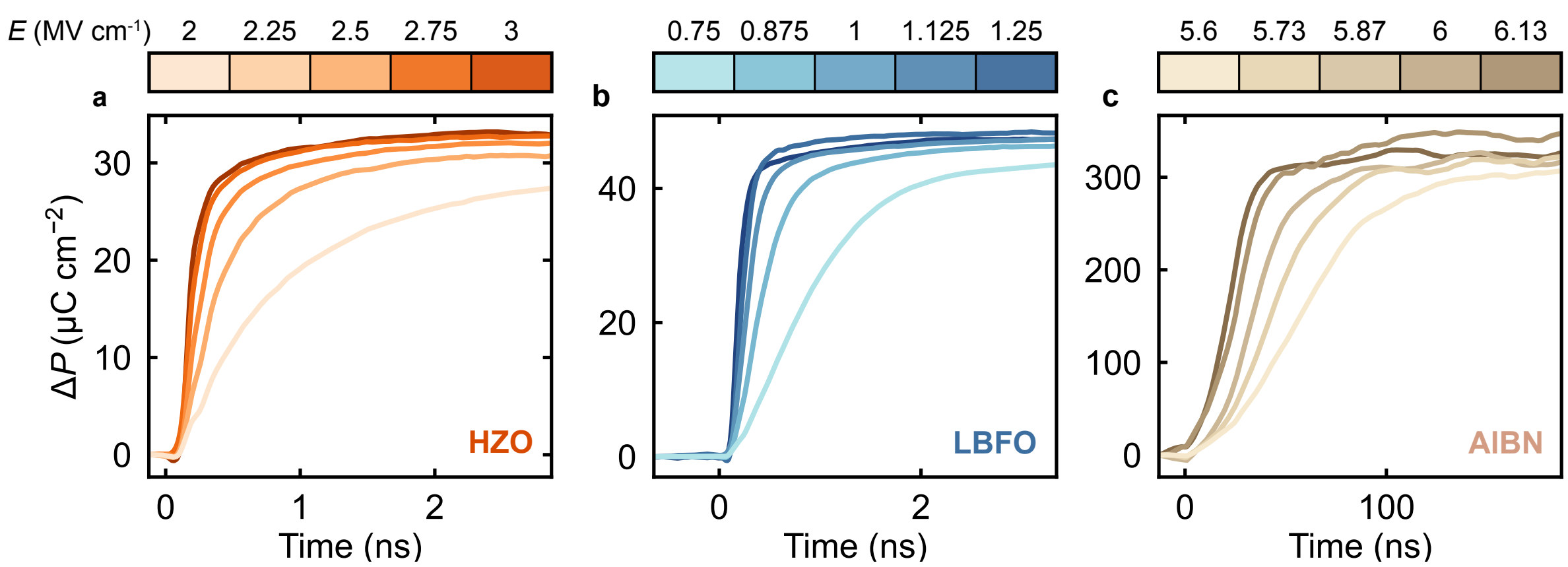

**Fig. S8 | Voltage-dependent polarization transients.** Polarization transients of 1 μm devices for HZO, LBFO, and AlBN (left to right). Unless otherwise specified, electric fields of 3, 1.125, and 6.13 MV cm$^{-1}$ were applied for HZO, LBFO, and AlBN, respectively, corresponding to 2.3 $V_c$ for HZO and LBFO.

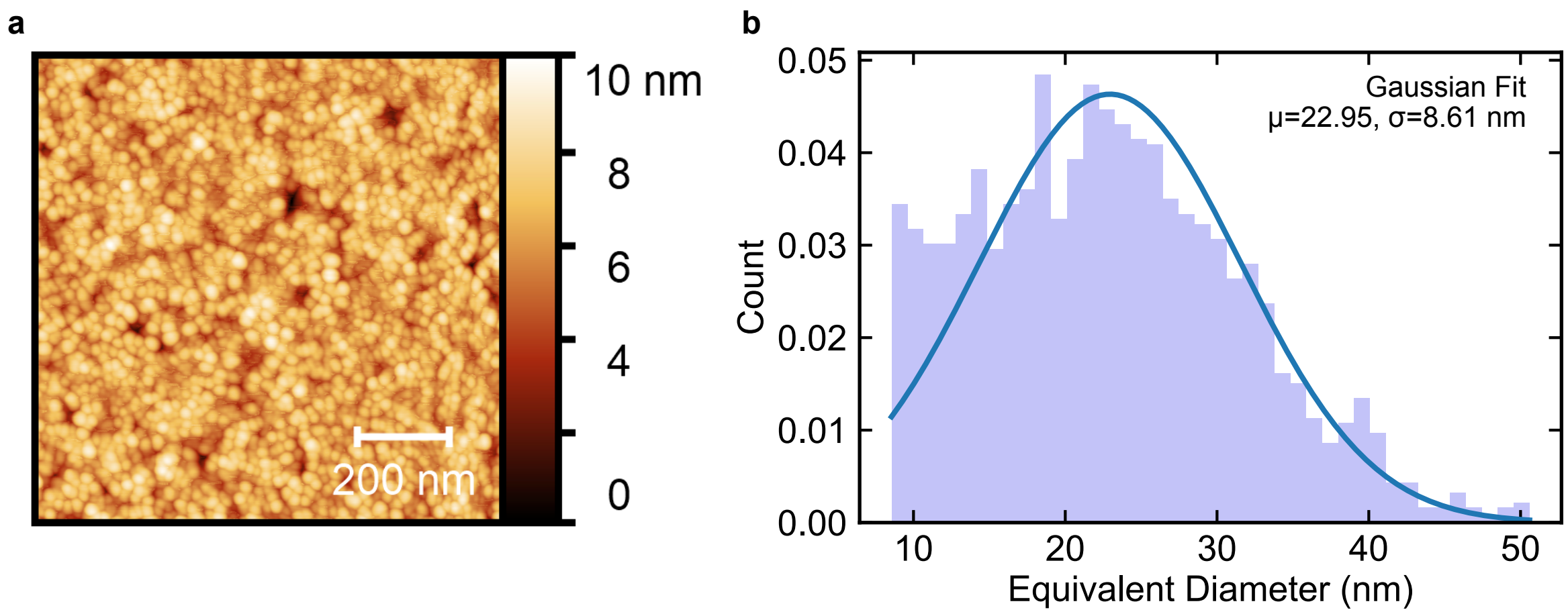

**Fig. S9 | Statistical analysis of the grain size in the polycrystalline HZO film. a,** AFM topography image of the 10 nm thick HZO film, revealing the presence of grains. **b,** Histogram of the grain size distribution, with solid line showing a fit to the Gaussian distribution, yielding a mean grain size of 22.95 nm.

## IV. Nucleation time

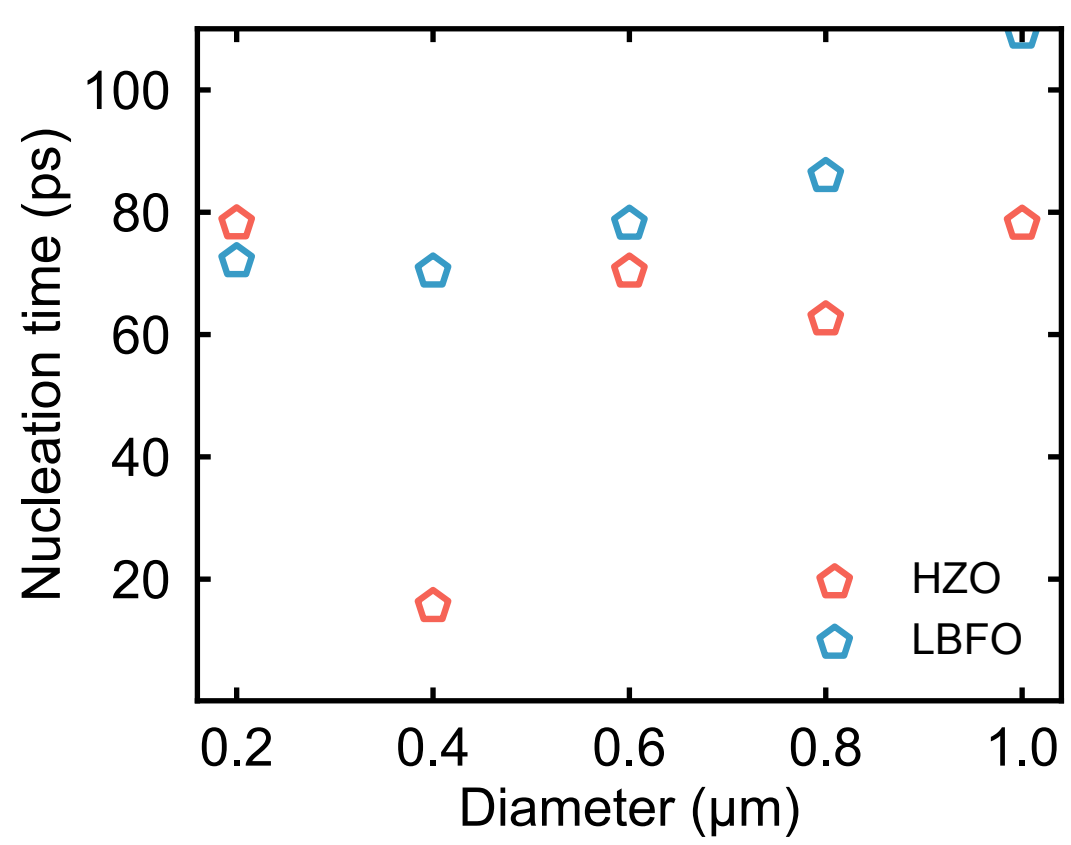

**Fig. S10 | Nucleation time.** The nucleation time for HZO (red) and LBFO (blue), defined as the interval time between the time point at which $V_{FE}$ reaches 43% (~1 $V_c$) of its maximum value and time point of 10% switched polarization.

The switching time in this study is defined as the time between 10% and 90% switched polarization, serving as a standardize metric for memory applications for consistent quantification. To capture the nucleation time, 0-10% switching times, for HZO are shown in Fig. S10. Here, 0% is defined as the time point at which $V_{FE}$ reaches 43% of its maximum value, corresponding to 1 $V_c$. A fast nucleation time, below 110 ps, is observed across all the device sizes shown here.